\documentclass[11pt,a4paper]{article}
\pdfoutput=1

\usepackage{jheppub}
\usepackage{rotating}

\usepackage{wrapfig}
\usepackage{amsfonts}
\usepackage{amsmath}
\usepackage{slashed}
\usepackage{amssymb}
\usepackage{latexsym}
\usepackage{multirow}
\usepackage{hyperref}
\usepackage{enumitem}
\usepackage{subfigure}
\hypersetup{colorlinks=true,citecolor=red,linkcolor=magenta,urlcolor=blue}
\usepackage{relsize}
\usepackage{booktabs}
\usepackage{cleveref}
\usepackage{fancyhdr}
\usepackage{float}
\usepackage{graphicx}
\graphicspath{ {./Figures/} }
\usepackage{microtype}
\usepackage{tabularx}
\usepackage{xcolor}
\usepackage[bottom]{footmisc}

\title{
EFT, decoupling, Higgs boson mixing, and higher dimensional operators
}

\abstract{
The effective field theory (EFT) framework is a precise approximation procedure
when the inherent assumptions of a large-scale separation between the Standard Model (SM) and new interactions alongside perturbativity are realised. Constraints from
available data might not automatically guarantee these circumstances when contrasted with UV scenarios that the EFT analysis wishes to inform. From an EFT perspective, achieving sufficient precision in navigating the alignment or decoupling limits beyond the SM scenarios can necessitate moving beyond the SM's leading, dimension six EFT deformation. Using the example of Higgs boson mixing, we demonstrated the importance of higher-dimensional terms in the EFT expansion. We analyse the relevance of virtual EFT corrections and dimension eight contributions for well-determined electroweak precision observables. We find that when moving away from the decoupling limit, the relevance of additional terms in the EFT expansion quickly becomes relevant. This demonstrates the necessity to move beyond dimension six interactions for any scenario that contains Higgs boson mixing.
}
\author[a]{Upalaparna Banerjee,}  
\author[a]{Joydeep Chakrabortty,}
\author[b]{Christoph Englert,} 
\author[b]{Wrishik Naskar,}
\author[a]{Shakeel Ur Rahaman,}
\author[c]{Michael Spannowsky}

\affiliation[a]{Indian Institute of Technology Kanpur, Kalyanpur, Kanpur 208016, Uttar Pradesh, India}
\affiliation[b]{School of Physics \& Astronomy, University of Glasgow, Glasgow G12 8QQ, United Kingdom}
\affiliation[c]{Institute for Particle Physics Phenomenology, Department of Physics, Durham University, Durham
	DH1 3LE, United Kingdom}

\emailAdd{upalab@iitk.ac.in} 
\emailAdd{joydeep@iitk.ac.in}
\emailAdd{christoph.englert@glasgow.ac.uk}
\emailAdd{w.naskar.1@research.gla.ac.uk}
\emailAdd{shakel@iitk.ac.in}
\emailAdd{michael.spannowsky@durham.ac.uk}

\preprint{IPPP/23/14}

\begin{document}
\maketitle

\flushbottom

\section{Introduction}
\label{sec:intro}

Effective field theory (EFT)~\cite{Weinberg:1978kz} is a formidable tool
for communicating sensitivity beyond the Standard Model (BSM) physics in times when particle physics data seemingly points towards a large-scale separation of new states relative to the Standard Model (SM) degrees of freedom. The extension of the SM by effective interactions relevant to the high-energy frontier of, e.g., the Large Hadron Collider (LHC), i.e. Standard Model Effective Theory (SMEFT) at dimension six~\cite{Grzadkowski:2010es} has received a lot of theoretical attention and improvement alongside its application in a series of experimental investigations. 
Matching calculations~\cite{Jiang:2018pbd, Haisch:2020ahr, Gherardi:2020det, Du:2022vso, Dawson:2017vgm, Zhang:2021jdf, Dittmaier:2021fls,Li:2022ipc, Zhang:2022osj,Carmona:2021xtq,Cohen:2020qvb,Bakshi:2018ics} that coarse grain ultra-violet (UV) BSM scenarios into EFT provide the technical framework to marry together concrete scenarios of new interactions with the generic EFT analysis of particle data. The latter is typically plagued with considerable uncertainties, both experimentally and theoretically. Even optimistic extrapolations of specific processes to the LHC's high luminosity (HL) phase can imply a significant tension between the intrinsic viability criteria that underpin the EFT limit setting trying to inform the UV scenarios' parameter spaces: EFT cut-offs need to be lowered into domains that can be directly resolved at the LHC. This can be at odds with the perturbativity of the obtained constraints (and hence limits the reliability of the fixed-order matching).

The obvious way out of this conundrum is to include higher-dimensional terms in the EFT expansion. Dimension eight interactions have increasingly moved into the focus of the theory community~\cite{Murphy:2020rsh, Li:2020gnx, Banerjee:2022thk}. From a practical point of view, this prompts the question of when we can be confident about reaching the point where phenomenologically-minded practitioners can stop. Unfortunately, an answer to this question is as process and model-dependent as matching a UV-ignorant EFT to a concrete UV scenario. 

Therefore, the phenomenological task is developing theory-guided intuition using representative scenarios that transparently capture key issues. The purpose of this note is to contribute to this evolving discussion using (custodial iso-singlet) Higgs boson mixing as an example. This scenario has seen much attention from the EFT perspective as the number of degrees of freedom and free parameters is relatively small, thus enabling a transparent connection of EFT and UV theory beyond the leading order of the EFT approach~(see, e.g., Refs.~\cite{Jiang:2018pbd,Dawson:2022cmu,Dawson:2021jcl}). Higgs mixing also arises in many BSM theories. We focus on electroweak precision observables as these are well-constrained by collider data, thus enabling us to navigate cut-offs and Wilson coefficients of the effective theory under experimental circumstances where precise predictions and matching are very relevant. This work is structured as follows: In Sec.~\ref{sec:elwprec}, we first discuss the oblique corrections and their relation to the polarisation functions to make this work self-contained; Sec.~\ref{subsec:RSFT} gives a quick discussion of the oblique corrections in the singlet scenario (see also~\cite{Pruna:2013bma,Binoth:1996au,Bowen:2007ia,Englert:2011yb,Batell:2011pz,Anisha:2021hgc}) with formulae provided in the appendix. We then focus on the oblique parameters for this case in dimensions six and eight SMEFT in Sec.~\ref{subsec:RSEFT}. We detail the comparison in Sec.~\ref{sec:fullvseft} with a view towards perturbative unitarity. Finally, we provide conclusions in Sec.~\ref{sec:conc}.
	
\section{Electroweak Precision Observables}
\label{sec:elwprec}
Extensions of the SM with modified Higgs sectors can be constrained through electroweak precision measurements. A famous subset of these that were instrumental in discovering the Higgs boson is the so-called oblique corrections parametrized by the Peskin-Takeuchi parameters \cite{PhysRevD.46.381} (see also~\cite{ALTARELLI1991161}). These $\widehat{S},\widehat{T},\widehat{U}$ are chiefly extracted from Drell-Yan-like production during the LEP era using global fits, e.g.,~Refs.~\cite{Bardin:1999yd,Flacher:2008zq}. 
Defining the off-shell two-point gauge boson functions for the SM gauge bosons as
\begin{equation}
\label{eq:twopoint}
\parbox{4.6cm}{\vspace{0.cm}\includegraphics[width=4.5cm]{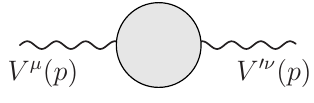}}
=\Pi_{VV'}(p^2)g^{\mu\nu}+\Sigma_{VV'}(p^2)p^\mu p^\nu,
\end{equation} 
with $V,V'=W,Z,\gamma$. The Peskin-Takeuchi parameters can then be written as
\begin{equation}
\begin{split}
\alpha \widehat{S}=&\left(\frac{4s_W^2c_W^2}{M_Z^2}\right)\left[\Pi_{ZZ}(M_Z^2)-\Pi_{ZZ}(0)-\Pi_{\gamma\gamma}(M_Z^2)-\frac{c_W^2-s_W^2}{c_Ws_W}\left(\Pi_{\gamma Z}(M_Z^2)-\Pi_{\gamma Z}(0)\right)\right],\\
\alpha \widehat{T}=&\frac{\Pi_{WW}(0)}{M_W^2}-\frac{\Pi_{ZZ}(0)}{M_Z^2}-\frac{2s_W}{c_W}\frac{\Pi_{\gamma Z}(0)}{M_Z^2},\\
\alpha \widehat{U}=& 4s_W^2\left[\frac{\Pi_{WW}(M_W^2)-\Pi_{WW}(0)}{M_W^2}-c_W^2\left(\frac{\Pi_{ZZ}(M_Z^2)-\Pi_{ZZ}(0)}{M_Z^2}\right)\right.\\
&\hspace{5cm}\left.-2s_Wc_W\left(\frac{\Pi_{\gamma Z}(M_Z^2)-\Pi_{\gamma Z}(0)}{M_Z^2}\right)-s_W^2\frac{\Pi_{\gamma\gamma}(M_Z^2)}{M_Z^2}\right]\,.
\end{split}
\end{equation}
$s_W,c_W$ denote the sine and cosine of the Weinberg angle, $\alpha$ is the fine structure constant, and $M_i$ stands for the gauge boson masses.\footnote{Note that we employ the normalization of Peskin and Takeuchi, although the hatted quantities are typically defined in the normalization of \cite{Barbieri:2004qk}. This is to avoid confusion between the oblique corrections and the singlet field introduced below.}
Constraints on new physics can then be formulated by examining the difference of these parameters from the best SM fit point. To draw a comparison between the full theory and EFT we restrict our analysis to one loop order. In the next subsection, we provide the contributions to the oblique parameters from the full theory.

\subsection{SM extended by a real singlet scalar}
\label{subsec:RSFT}
The extension of the Standard Model by a real singlet scalar finds its relevance in various compelling instances, such as the electroweak hierarchy problem and dark matter~\cite{Curtin:2015bka,Craig:2013xia,Curtin:2014jma,Espinosa:2011ax,Barger:2011vm,Profumo:2007wc,Huber:2006wf,Menon:2004wv,Mambrini:2011ik,Gonderinger:2009jp,He:2008qm,Burgess:2000yq,McDonald:1993ex,SILVEIRA1985136}. Additionally, extensive research has been conducted at the LHC~\cite{Datta:1997fx,Bahat-Treidel:2006zof,Barger:2006sk,OConnell:2006rsp,Barger:2007im,Barger:2008jx} to analyse the distinctive features of the singlet scalar, particularly its influence on the characteristics of the Higgs. Moreover, the inherent simplicity of this model serves as an additional impetus for delving into precision and matching calculations, and this has made it a focal point in recent EFT studies~\cite{Henning:2014wua,delAguila:2016zcb,deBlas:2014mba,Gorbahn:2015gxa,Brehmer:2015rna,Chiang:2015ura,Buchalla:2016bse,Jiang:2016czg,Corbett:2017ieo,Ellis:2023zim}. Building on this motivation, we begin our analysis by describing a scalar potential for the SM Higgs Sector extended by a real singlet scalar field ($S$),
\begin{equation}
\label{eqn:potential}
\begin{split}
    V(H,S)=-\mu^2 H^\dagger H+\frac{1}{2}m^{2}_{S}&S^2+\eta_S\left(H^\dagger H-\frac{v^2}{2}\right) S \\
    &+k_S\left(H^\dagger H -\frac{v^2}{2}\right) S^2+\lambda_h(H^\dagger H)^2+\frac{1}{4!}\lambda_SS^4,
\end{split}
\end{equation}
with $H$ being the SM Higgs doublet, which gets a vacuum expectation value (vev) $v\simeq 246~\text{GeV}$.  
$H$ can then be expanded around the vev: 
\begin{equation}
    H=\frac{1}{\sqrt{2}}\begin{pmatrix}
    \sqrt{2}\,G^{+}\\
    v+h+i\,\eta
    \end{pmatrix}.
\end{equation}
The minimisation of the potential relates the Lagrangian parameters $\lambda_h$, $\mu$ and the Higgs vev, $\mu^2=v^2\lambda_h$. 
The presence of the mixing terms in the potential given in Eq.~\eqref{eqn:potential}, results in different mass eigenstates, corresponding to the mass mixing matrix
given by,

\begin{equation}
    (M^2)_{ij}=\begin{pmatrix}
        2v^2\lambda_h& &v\eta_S\\
        v\eta_S& &m_S^2
    \end{pmatrix},
\end{equation}
that are a mixture of $h$ that is the neutral component of $H$, and $S$, related by the mixing angle $\theta$:   
\begin{equation}
\label{eqn:mixing}
    \begin{pmatrix}
        \tilde{h}\\
        \tilde{s}
    \end{pmatrix}=\begin{pmatrix}
        \cos{\theta}& &-\sin{\theta}\\
        \sin{\theta}& &\cos{\theta}
    \end{pmatrix}\begin{pmatrix}
    h\\
    S
    \end{pmatrix}.
\end{equation}
The mass eigenvalues corresponding to the physical parameters shown in correspond to the masses of the scalars in the theory, i.e., $M_{h}=125~\text{GeV}$ and a free parameter $M_{\cal{S}}$, respectively. These mass eigenvalues can be expressed in terms of the Lagrangian parameters and the Higgs vev ($v$) as,
\begin{eqnarray}
\label{eq:mass}
    M_h^2,M_\mathcal{S}^2=\frac{1}{2}\left(m_S^2+m_h^2\mp\sqrt{4v^2\eta_S^2+(m_S^2-m_h^2)^2}\right)\, ,
\end{eqnarray}
where $m_h^2=2\lambda_h v^2=2\mu^2$.
The mixing angle $\cos{\theta}$ can be written in terms of the physical pole masses and the parameter $\eta_S$ as
    \begin{equation}
    \label{eq:cos2theta}
        \cos^2{\theta} = \frac{1}{2} \left[ 1+
{[(M_{\cal{S}}^2-M_h^2)^2-4\eta_S^2 v^2]^{1/2}\over M_{\cal{S}}^2-M_h^2} \right] \,.
    \end{equation}
This  implies $\cos\theta = 1$ for $\eta_S=0$. We will focus on the region of large $M_{\cal{S}}\gg M_h$ for our EFT comparison; in this region
\begin{equation}
\label{eq:expand}
\cos^2\theta \simeq
1 - {\eta_S^2 v^2\over M_{\cal{S}}^4} \,.
\end{equation}
Eqs.~\eqref{eq:mass} and~\eqref{eq:cos2theta} can be used to relate the Lagrange parameters $m_{S,h}^2$ to the pole masses.

\begin{figure}[!t]
     \centering
 \includegraphics[width=0.7\textwidth]{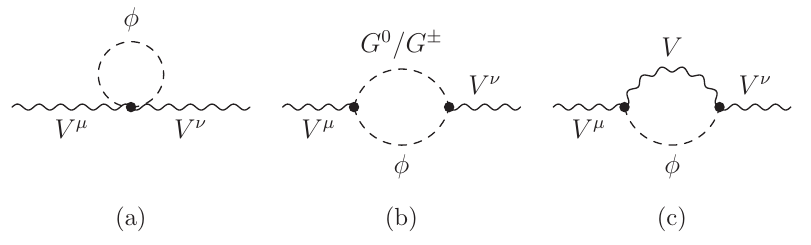}
     \caption{Relevant Feynman diagrams with scalars in the loop that have been considered to compute the oblique corrections. Here $\phi\in (\tilde{h}, \tilde{s})$ when one-loop correction in the full theory is computed.}
     \label{fig:feynman_diagrams}
\end{figure}

For the computation of the oblique parameters we only consider the radiative corrections from the scalar-involved diagrams shown in Fig.~\ref{fig:feynman_diagrams}, since the other diagrams will provide the same contribution to BSM and SM theory, therefore dropping out from the deviation. The explicit expressions are given in the appendix~\ref{app:fullsing}.

\begin{figure}[!htpb]
\centering
\includegraphics[width=0.55\linewidth]{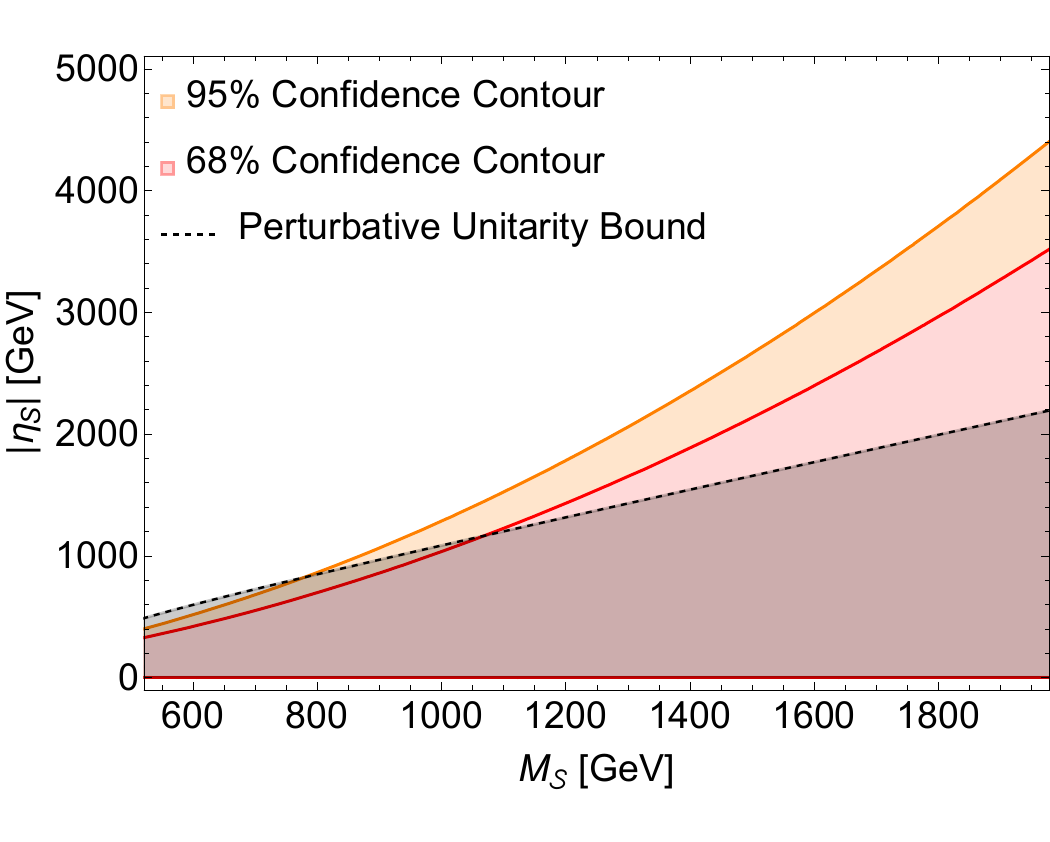}
\caption{$95\%$ and $68\%$ confidence interval bounds on the BSM trilinear coupling $\eta_S$ and the Mass of the heavy scalar ($M_{\cal{S}}$) obtained from the full-theory calculation setting constraints from {\sc{GFitter}}~\cite{Haller:2018nnx}. An additional unitarity bound is superimposed which identifies a region of $\eta_S$ where perturbative matching is applicable
(see appendix~\ref{sec:unitarity} for details), this region is limited by the approximate bound $\eta_S \approx M_{\mathcal{S}}$.}
\label{fig:full_theory_CI}
\end{figure}

\allowdisplaybreaks
Eq.~\eqref{eqn:mixing} clearly shows that the light (heavy) scalar couplings to the SM particles are suppressed by a factor of $\cos\theta$ ($\sin\theta$). Therefore, the contributions to the gauge boson self-energies get modified by a factor of $\cos^2{\theta}$ or $\sin^2\theta$ depending on the neutral scalar coupled to. 
We then express the mixing angle regarding the BSM parameters in the potential. 
Limits are then imposed on the independent BSM parameters (in our case, we choose this $\eta_S$, the mixing angle can then be inferred from Eq.~\eqref{eq:cos2theta} for allowed regions) and the mass of the heavy scalar ($M_{\cal{S}}$) using the constraints of {\sc{GFitter}} data~\cite{Haller:2018nnx} as shown in Fig.~\ref{fig:full_theory_CI}.

\subsection{Real Singlet Model from SMEFT perspective}
\label{subsec:RSEFT}
\begin{wrapfigure}[8]{l}{6cm}
     \parbox{5.5cm}{
     \centering
     \includegraphics[scale=0.5]{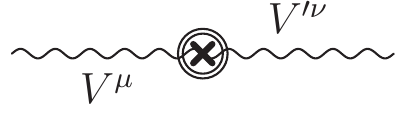}
     \caption{Tree-level correction to the gauge boson propagators due to the presence of effective operators.}
     \label{fig:tree-eft-correction}}
\end{wrapfigure}

To investigate how well the effective theory replicates the minute signatures of the singlet extension of the SM described in Sec.~\ref{subsec:RSFT} or, in turn, adjudge the significance of the higher-order effective corrections, we extend the effective series with relevant operator structures till dimension eight:
\begin{equation}
{\cal{L}} =
{\cal{L}}_{\text{SM}} + \sum_i \frac{{\cal{C}}_i^{(6)}}{ \Lambda^2} \,{\cal{O}}_i^{(6)} + \sum_j \frac{{\cal{C}}_j^{(8)}}{\Lambda^4} \,{\cal{O}}_j^{(8)}.
\end{equation}
Here, the Wilson coefficients $\mathcal{C}_{i}$ parametrize the strengths of the operators $\mathcal{O}_{i}$ that are produced after integrating out the heavy real singlet scalar (for a complete matching of such operators at dimension six, see Refs.~\cite{Banerjee:2022thk}). We have chosen the cut-off scale $\Lambda$ to be $m_S$. This implies that in the case of effective theory, there may be tree-level electroweak corrections, as shown in Fig.~\ref{fig:tree-eft-correction}, to  Eq.~\eqref{eq:twopoint} from the effective operators that
may emerge in the process of integrating out heavy fields from the UV diagram and(or) through the renormalization group running of effective operators generated by integrating out tree-level heavy propagator at the cut-off scale.  These contributions depend on the renormalization scale and play an essential role in our further computation, see also~\cite{Jiang:2018pbd}.
Depending on whether the operators that could contribute to the dominant (when considered in a model-independent way) tree-level correction, as shown in Fig.~\ref{fig:tree-eft-correction}, are generated at one-loop itself, the contributions from the operators generated at the tree level, which can modify the interactions at the one-loop, can become significant.
 
We categorically list the effective corrections to $\widehat{S}, \widehat{T} $ up to one loop:
\begin{itemize} 
    \item \underline{\textbf{Tree-level correction}}: Expanding the Lagrangian with dimension six and dimension eight operators can
induce corrections to the transverse tree-level vector boson propagators ($\Pi_{VV'}$) itself, which in turn modifies $\widehat{S}$, $\widehat{T}$ parameters~\cite{Murphy:2020rsh}
\begin{equation}
    \begin{split}
    \label{Eq:tree-correction}
         \widehat{S}_{\text{eft,tree}} &= \frac{4\, c_W \,s_W \,v^2}{\alpha}
         \,\mathcal{C}_{HWB}^{(6)}\,+\frac{2\, c_W \,s_W \,v^4}{\alpha}
         \,\mathcal{C}_{HWB}^{(8)}
        ,\\
         \widehat{T}_{\text{eft,tree}} &= -\frac{v^2}{2\,\alpha}\,\mathcal{C}_{H\mathcal{D}}^{(6)}\,-\frac{v^4}{2\,\alpha}\mathcal{C}_{H\mathcal{D},2}^{(8)}.
    \end{split}
     \end{equation}
 The expressions for modifying the individual $\Pi_{VV'}$ functions are given in the appendix~\ref{app:pi-tree-eft}. The dimension six operators contributing to Eq.~\eqref{Eq:tree-correction} are generated at one-loop while integrating out the heavy field. The matching expressions for these coefficients are given in Tab.~\ref{Tab:relevant_ops_d6}. We have also computed the one-loop matching for the dimension eight operators involved in Eq.~\eqref{Eq:tree-correction} and noticed that these do not receive any correction while integrating out complete heavy loop diagrams. On the other hand, these coefficients receive contributions from removing the redundant structures at dimension six, as discussed in Ref.~\cite{Banerjee:2022thk}. Since the latter corresponds to a two-loop suppressed sub-leading contribution, we neglect the associated effects in our analysis.
\begin{table}[!t]
	\centering \scriptsize
	\renewcommand{\arraystretch}{2.2}
     \resizebox{0.67\textwidth}{!}{
\begin{tabular}{|c|c|c|}
		\hline
		\textsf{\quad Operator}$\quad$&
		\textsf{\quad Op. Structure}$\quad$&
            \textsf{\quad Wilson coeffs.}$\quad$\\
		\hline
                        
            $\mathcal{O}_{H\square}^{(6)}$&
            $(H^{\dagger}H)\square(H^{\dagger}H)$&
            $-\frac{\eta_{\mathcal{S}}^2}{2\,m_{S}^4}$\\
            \hline

            $\textcolor{blue}{\mathcal{O}_{HWB}^{(6)}}$&
            $\textcolor{blue}{(H^{\dagger}\tau^{I}H)W^{I}_{\mu\nu}B^{\mu\nu}}$&
            $\textcolor{blue}{\frac{g_{_W}g_{Y}\eta_{\mathcal{S}}^2}{128\,m_{S}^4\,\pi^2}}$\\
            \hline 
            
            $\textcolor{blue}{\mathcal{O}_{HD}^{(6)}}$&
            $\textcolor{blue}{(H^{\dagger}\mathcal{D}_{\mu}H)^{*}(H^{\dagger}\mathcal{D}^{\mu}H)}$&
            $\textcolor{blue}{-\frac{7g_{Y}^2\eta_{\mathcal{S}}^2}{288\,m_{S}^4\,\pi^2}}$\\
            \hline

	\end{tabular}
 }
	\caption{\small Relevant operators that produce 
        tree and one-loop corrections to the gauge boson self-energy. The structures in blue first appear at tree-level correction, whereas the rest of the operators contribute at one-loop first. }
	\label{Tab:relevant_ops_d6}
\end{table}
    \item \underline{\textbf{One-loop insertion of operators}}: One-loop corrections to the oblique parameters are essential for the tree-level generated operators, for they provide a similar contribution as the operators that are produced at one-loop
    contributing to the tree-level propagator corrections shown in Eq.~\eqref{Eq:tree-correction} for a model-dependent analysis. In our case, such a contribution arises from the operators $\mathcal{O}_{H\square}^{(6)}$ and $\mathcal{O}_{H\mathcal{D},1}^{(8)}$. The explicit forms of their structures are given in Tabs.~\ref{Tab:relevant_ops_d6} and~\ref{Tab:relevant_ops_d8}, respectively. These operators modify the canonical form of the kinetic term for the Higgs field
    \begin{eqnarray}
        \mathcal{L}_{h,\text{kin}} &=& \frac{1}{2}\,\left(1-2\,v^2\mathcal{C}_{H\square}^{(6)}+\frac{v^4}{4}\mathcal{C}_{H\mathcal{D},1}^{(8)}\right)(\partial_{\mu}h)^2,
    \end{eqnarray}
    which can be removed by redefining the field $h' \to Z_{h} \,h$ with 
    \begin{eqnarray}\label{eq:zh}
        Z_{h} = \left(1-v^2\,\mathcal{C}_{H\square}^{(6)}+\frac{v^4}{8}\,\mathcal{C}_{H\mathcal{D},1}^{(8)}\right).
    \end{eqnarray}
    This implies that while computing the higher order corrections for EFT, we need to recall that $\phi=h' \to Z_h \,h$ in Fig.~\ref{fig:feynman_diagrams}. This also accounts for suitable modifications in the vertices, involving  Higgs and Goldstone in Fig.~\ref{fig:feynman_diagrams}. 
    This correction, up to $\mathcal{O}(1/\Lambda^4)$ capturing the effects from both $(\text{dimension six})^2$ and dimension eight terms,  is incorporated by replacing the $ \cos^2{\theta} $ with $Z_h^2$ and setting $ \sin^2\theta $ to zero in the expressions shown in appendix~\ref{app:fullsing}.

     \item \underline{\textbf{RGE improved correction}}: It is important to include the running effects to the Wilson coefficients which arise at tree-level. $\widehat{T}_{\text{eft,tree}}$ in Eq.~\eqref{Eq:tree-correction} at dimension six receives such an additional contribution from the operator $\mathcal{O}_{H\mathcal{D}}^{(6)}$. Contributions arising from the running of the coefficient of the operator $\mathcal{O}_{H\square}^{(6)}$~\cite{Jenkins:2013zja,Alonso:2013hga}
     \begin{eqnarray}
     \label{eq:RGE-corrction-chd6}
         16 \pi^2 \,\frac{d\,\mathcal{C}^{(6)}_{H\mathcal{D}}\,(\mu)}{d\ln{\mu}} & = & \frac{20}{3}\,g_{Y}^2\,\mathcal{C}_{H\square}^{(6)}, \nonumber\\
       \Longrightarrow \, \Delta\,\mathcal{C}_{H\mathcal{D}}^{(6)}\,|_{\text{RGE}} & = & -\frac{5\,g_{Y}^2\eta_{\mathcal{S}}^2}{24\,\pi^2\,m_{S}^4}\,\log{\left[\frac{M_Z}{m_S}\right]}.
    \end{eqnarray}
    The total contribution to $\mathcal{C}_{H\mathcal{D}}$ at the EW scale becomes
    \begin{eqnarray}\label{eq:total-corrction-chd6}
        \mathcal{C}_{H\mathcal{D}}^{(6)} (M_Z) = {-\frac{7g_{Y}^2\eta_{\mathcal{S}}^2}{288\,\pi^2\,m_S^4}} -\frac{5\,g_{Y}^2\eta_{\mathcal{S}}^2}{24\,\pi^2\,m_S^4}\,\log{\left[\frac{M_Z}{m_S}\right]}\;.\nonumber\\
    \end{eqnarray}
     The part of the beta function (cf.~Eq.~\eqref{eq:RGE-corrction-chd6}) for the dimension eight Wilson coefficient $\mathcal{C}_{H\mathcal{D},2}^{(8)}$ and $\mathcal{C}_{HWB}^{(8)}$ stems from 
     \begin{eqnarray}\label{eq:d8-rge}
         16 \pi^2 \,\frac{d\,\mathcal{C}_{H\mathcal{D},2}^{(8)}\,(\mu)}{d\ln{\mu}}&=& \frac{40}{3}\,g_{Y}^2\,(\mathcal{C}_{H\square}^{(6)})^2\, +\cfrac{10}{3}g_Y^2  \mathcal{C}_{H\mathcal{D},1}^{(8)}\, + \mathcal{C}_{H^4D^4,3}^{(8)}\left(-\cfrac{11}{24} g_Y^4 -\cfrac{79}{48}g_Y^2 g_W^2 + 3 g_Y^2 \lambda_h\right) \nonumber\\
         \Longrightarrow \, \Delta\,\mathcal{C}_{H\mathcal{D},2}^{(8)}\,|_{\text{RGE}} & = & \cfrac{1}{16 \pi^2}  \Bigg[ \frac{10\,g_{Y}^2\eta_{\mathcal{S}}^4}{3\,m_S^8} +\cfrac{10}{3}g_Y^2 \left( \frac{4\eta_{\mathcal{S}}^2k_{\mathcal{S}}}{m_S^6}-\frac{8\lambda_{\mathcal{S}}\eta_{\mathcal{S}}^2k_{\mathcal{S}}}{m_S^6} \right) \nonumber \\
         & &  + \cfrac{2\eta_\mathcal{S}^2}{m_S^6}  \left( -\cfrac{11}{24} g_Y^4-\cfrac{79}{48}g_Y^2 g_W^2 + 3 g_Y^2 \lambda_h  \right) \Bigg] \log{\left[\frac{M_Z}{m_S}\right]}, \nonumber \\
         16 \pi^2 \,\frac{d\,\mathcal{C}_{HWB}^{(8)}\,(\mu)}{d\ln{\mu}}&=&  \mathcal{C}_{H^4D^4,3}^{(8)}\left(-\cfrac{11}{48} g_Y^3 g_W -\cfrac{29}{24}g_Y g_W^3 + g_Y g_W \lambda_h\right)  \nonumber \\
         \Longrightarrow \,\Delta\,\mathcal{C}_{HWB}^{(8)}\,|_{\text{RGE}} & = &  \cfrac{\eta_\mathcal{S}^2}{8 \pi^2 m_S^6} \left(-\cfrac{11}{48} g_Y^3 g_W -\cfrac{29}{24}g_Y g_W^3 + g_Y g_W \lambda_h\right) \log{\left[\frac{M_Z}{m_S}\right]}.
     \end{eqnarray} 
\end{itemize}   
with $g_Y=e/c_W, g_W=e/s_W$.  Since we aim to make a comparative analysis between the full theory and the EFT predictions for the oblique parameters, it is necessary to express both of them in terms of the same model parameters. To achieve this, once we find the relevant WCs in terms of $m_S$, we must substitute it with the physical mass ($M_\mathcal{S}$). In the framework of EFT, this substitution needs to be performed carefully, given that $M_\mathcal{S}^2 = m_S^2+ \Delta_{\text{NLO}}$, where $\Delta_{\text{NLO}} \approx v^2\eta_S^2/m_S^2$.The next order term in this expansion is $\approx -v^4\eta_S^4/m_S^6$, i.e. the $m_S$ suppression in this order is equivalent to dimension ten corrections, hence we neglect this term, and any higher order contribution beyond this. We express $m_S$ in terms of $M_\mathcal{S}$ implementing the correction stemming from $\Delta_{\text{NLO}}$ to use the WCs for further analysis.  In addition to contributions from dimension six effective operators, we also compute the contribution to dimension eight operators from the equations of motion of the dimension six operators and the  RGE-improved corrections due to dimension six operators, see further Refs.~\cite{Li:2020gnx,Murphy:2020rsh,Chala:2021pll,DasBakshi:2022mwk}. 

\section{Full theory vs EFT}
\label{sec:fullvseft}
In this section, we compare a full theory and its effective version captured in SMEFT. We carefully investigate how the inclusion of higher mass dimensional operators, suppressed by the mass of the heavy integrated-out field, in the EFT expansion, affects the computation of our chosen observable $(\widehat{T}, \widehat{S})$. For this, we categorize the EFT contribution into two parts. To start with, we discuss the dimension six $(d_6)$ part, containing linear dimension six Wilson coefficients (WCs). Here, we include the cumulative effects of field redefinition and radiative corrections on the oblique parameters. Then, we consider the corrections from the dimension eight $(d_8)$ operators that are linear functions of dimension eight WCs. We also include the dimension eight equivalent contributions (referred to as $(d_6)^2$), from dimension six operators which are quadratic functions of dimension six WCs. This takes care of the radiative generation of dimension eight operators from dimension six ones, see Eq.~\eqref{eq:d8-rge} and the expansion of $Z_h^2$. We list all those operators that contribute in different orders:
\begin{itemize}
    \item $d_6$:\; $\mathcal{C}_{HD}^{(6)},\; \mathcal{C}_{HWB}^{(6)},\; \mathcal{C}_{H\square}^{(6)}$\,;  

    \item  $d_8$:\; $\mathcal{C}_{H\mathcal{D},1}^{(8)},\; \mathcal{C}_{H\mathcal{D},2}^{(8)},\;  \mathcal{C}_{H^4D^4,3}^{(8)},\; \mathcal{C}_{HWB}^{(8)}$\,;

    \item  $(d_6)^2:$ $(\mathcal{C}_{H\square}^{(6)})^2$\, . 
\end{itemize}

\begin{table}[!t]
	\centering \scriptsize
	\renewcommand{\arraystretch}{2.2}
     \resizebox{0.67\textwidth}{!}{
\begin{tabular}{|c|c|c|}
		\hline
		\textsf{\quad Operator}$\quad$&
		\textsf{\quad Op. Structure}$\quad$&
            \textsf{\quad Wilson coeffs.}$\quad$\\
		\hline
                        
            $\mathcal{O}_{HD,1}^{(8)}$&
            $(H^{\dagger}H)^2\,(\mathcal{D}_{\mu}H^{\dagger}\mathcal{D}_{\mu}H)$&
            $\frac{4\eta_{\mathcal{S}}^2k_{\mathcal{S}}}{m_S^6}-\frac{8\lambda_{\mathcal{S}}\eta_{\mathcal{S}}^2k_{\mathcal{S}}}{m_S^6}$\\
            \hline

            $\mathcal{O}_{H^4D^4,3}^{(8)}$&
            $(\mathcal{D}_{\mu}H^{\dagger}\mathcal{D}^{\mu}H)(\mathcal{D}_{\nu}H^{\dagger}\mathcal{D}^{\nu}H)$&
            $\frac{2\eta_\mathcal{S}^2}{m_S^6}$\\
            \hline
            
	\end{tabular}
 }
	\caption{\small Relevant dimension eight operators that produce one-loop equivalent corrections to the gauge boson self-energy. These operators first appear while integrating out heavy fields at tree-level.} 
	\label{Tab:relevant_ops_d8}
\end{table}

\begin{figure}[!t]
     \centering
     \subfigure[\label{fig:eta_vs_M}]{\includegraphics[width=0.495\textwidth]{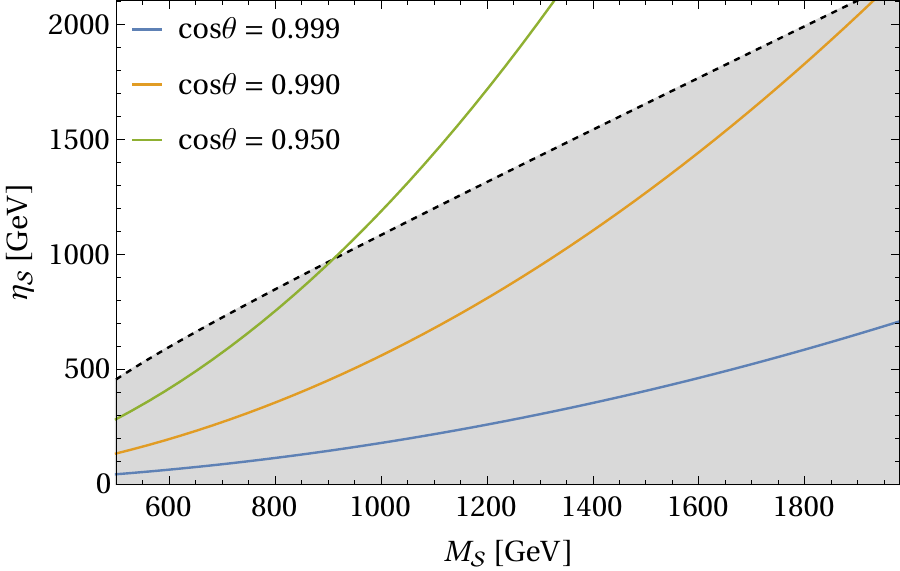}}\hfill  
     \subfigure[\label{fig:costheta_vs_M}]{\includegraphics[width=0.495\textwidth]{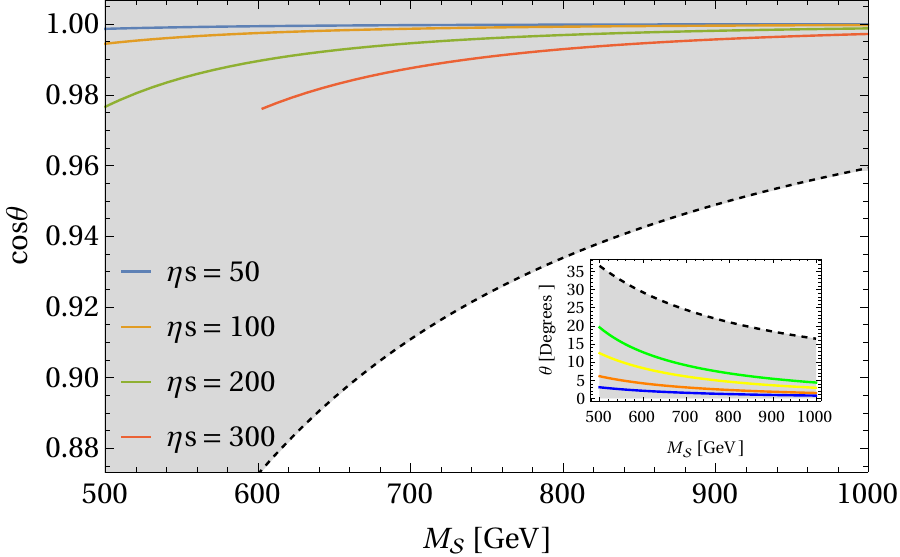}}
     \caption{(a) shows the variation of the trilinear coupling with respect to heavy scalar mass. In (b) we show how the mixing angle varies as a function of the heavy mass for fixed values of the trilinear coupling $\eta_\mathcal{S}$. In both plots, the gray-shaded region respects perturbative unitarity (see appendix~\ref{sec:unitarity}). The first plot indicates that when moving away from the traditional decoupling limit ($\cos \theta \sim 1$), a small portion of the $M_\mathcal{S}-\eta_{\mathcal{S}}$ plane is permitted by unitarity. However, the region of $M_\mathcal{S}-\eta_{\mathcal{S}}$ that corresponds to $\cos \theta \sim 1$ is always allowed. The second plot shows how a large value of $\eta_\mathcal{S}$ can shift the $\cos\theta$ value away from 1 while still maintaining unitarity.}
     \label{fig:constant_mixing_angle}
\end{figure}
\begin{figure}[!t]
     \centering
     \includegraphics[width=0.70\linewidth]{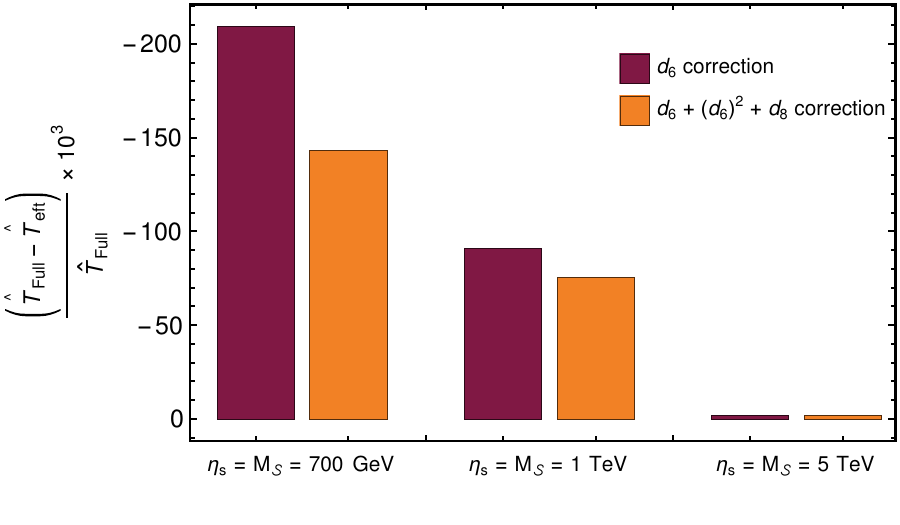}
     \caption{Impact of individual contributions on relative $\Delta \widehat{T}=(\widehat{T}_{\text{Full}}-\widehat{T}_{\text{eft}})/\widehat{T}_{\text{Full}}$ for three benchmark choices of $\eta_s=M_S$, having nearly-maximal allowed mixing. The total contributions up to dimension eight effective operators including the running effects reduce $\Delta \widehat{T}$ signifying inclusion of dimension eight operators brings the effective theory prediction relatively closer to that from the full theory. Comparing the outcome corresponding to each benchmark point, it becomes evident that for the lower mass range of the heavy field, the correction from dimension eight produces a more pronounced impact. }
     \label{fig:comparison_of_different_terms_deltaT}
\end{figure}
\begin{figure}[!t]
     \centering
     \includegraphics[width=0.70\linewidth]{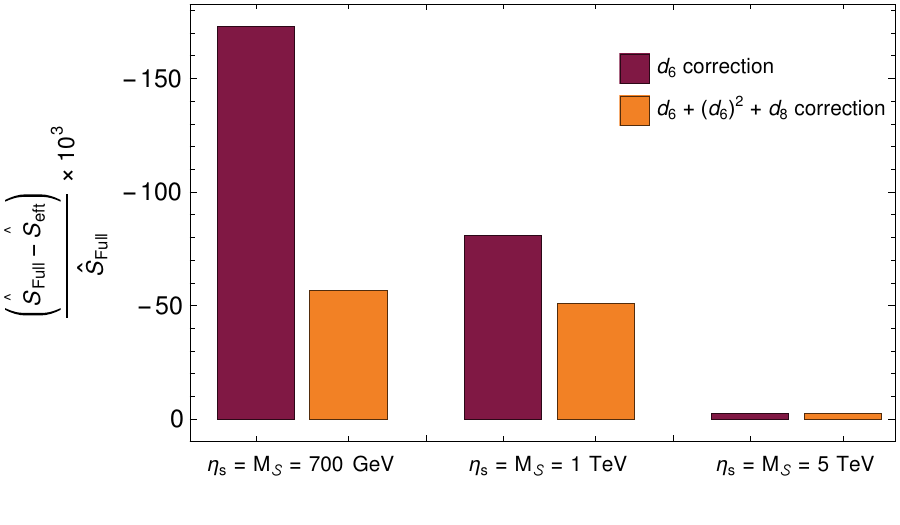}
     \caption{Impact of individual contributions on relative $\Delta \widehat{S}=(\widehat{S}_{\text{Full}}-\widehat{S}_{\text{eft}})/\widehat{S}_{\text{Full}}$ for three benchmark choices of $\eta_s=M_S$. Similar to the case of $\Delta \widehat{T}$, corrections up to dimension eight including the running effects leave more impact when the heavy field is relatively lighter, even though, they do not have the same magnitude as seen in the case of $\Delta \widehat{T}$. This makes the truncation of the effective series a process-dependent (i.e. choice of observable) issue.} 
     \label{fig:comparison_of_different_terms_deltaS}
\end{figure}
 \begin{figure}[!t]
     \centering
     \subfigure[~]{\includegraphics[width=0.48\textwidth]{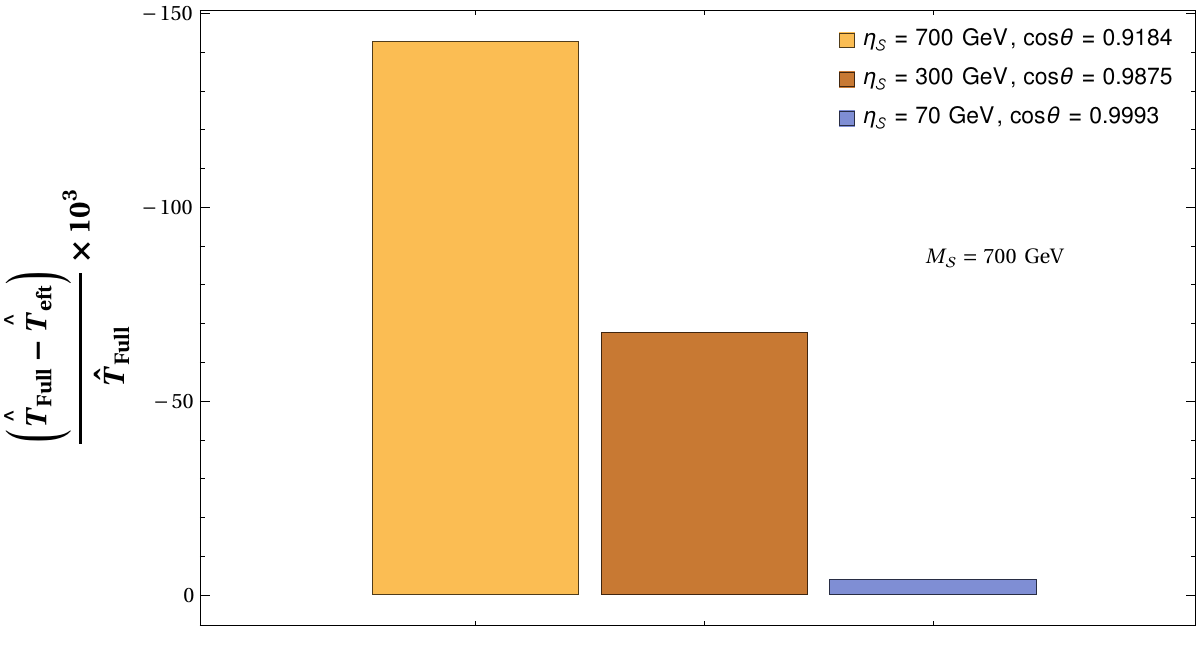}}
     \hfill
     \subfigure[~]{\includegraphics[width=0.48\textwidth]{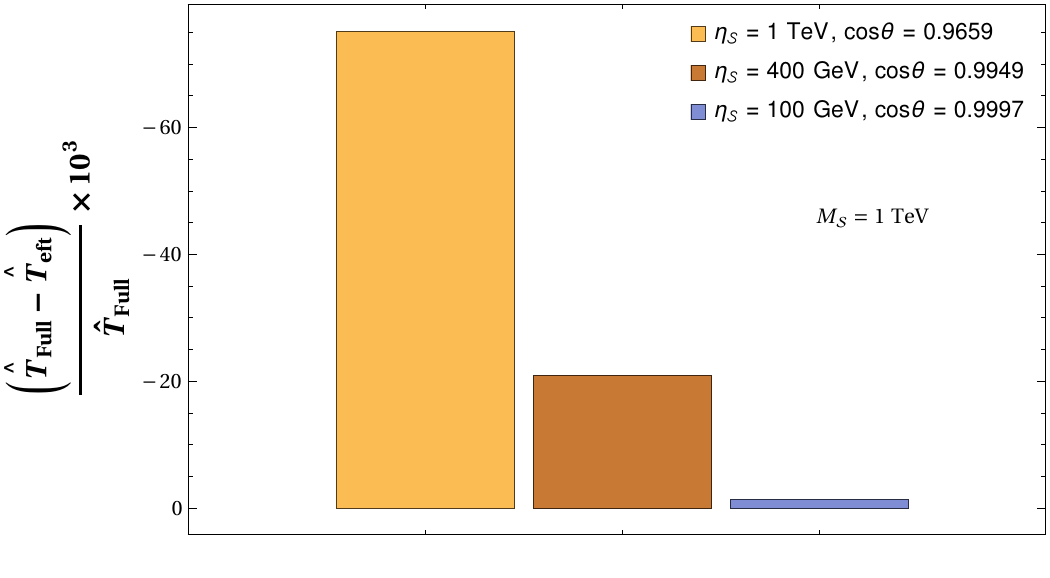}}
     \hfill
     \subfigure[~]{\includegraphics[width=0.48\textwidth]{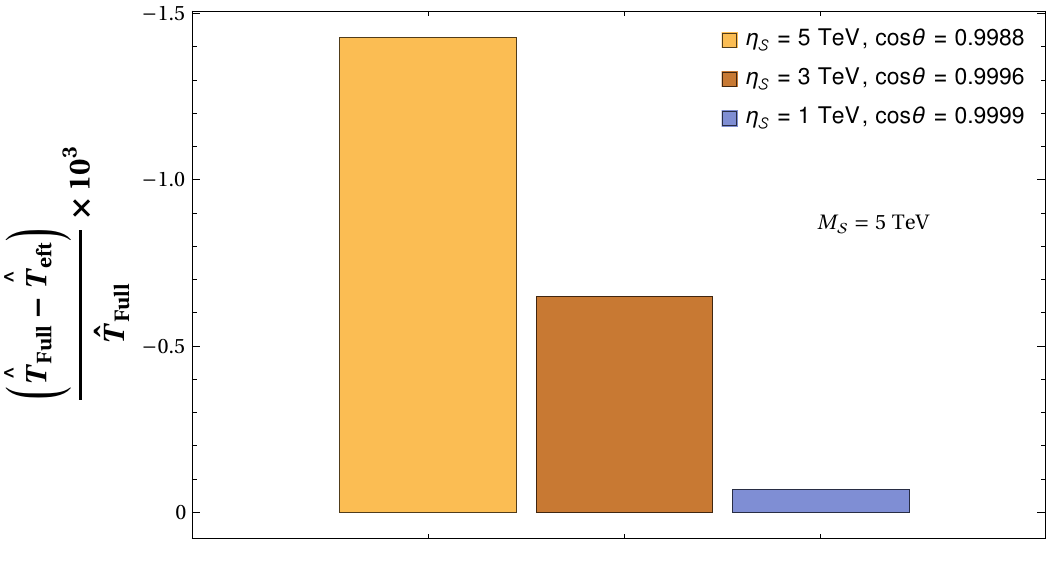}}\hfill
     \parbox{0.48\textwidth}{
     \vspace{-2cm}\caption{Relative difference between the full theory and the EFT  computation for $\widehat{T}$ parameter at different heavy field mass scales. In this case, we have performed the EFT expansion to accommodate contributions from up to dimension eight effective operators including the running of all contributory operators as well. The mass scales are chosen to be (a) 700 GeV, (b) 1 TeV, and (c) 5 TeV. The values for $\eta_S$ are chosen such that they satisfy the unitarity bounds.}
     \label{fig:deltaT_for_constant_eta}}
\end{figure}
 \begin{figure}[!t]
     \centering
     \subfigure[~]{\includegraphics[width=0.48\textwidth]{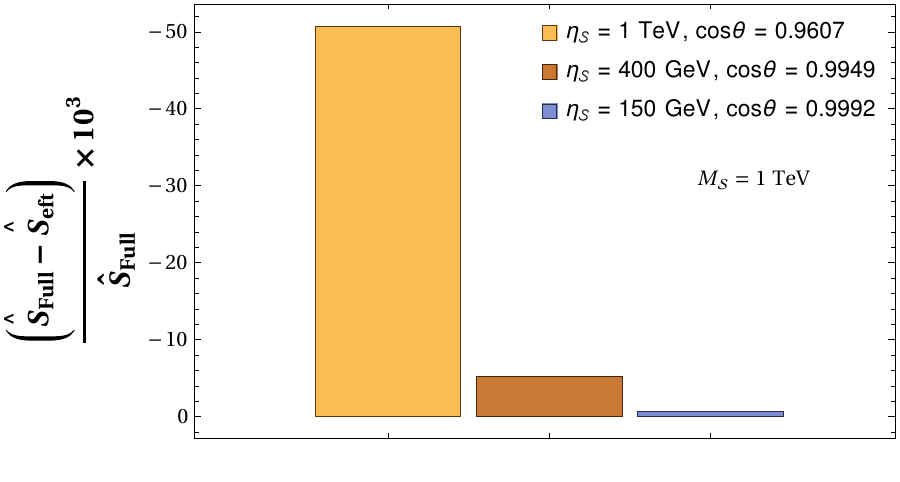}}
     \hfill
     \subfigure[~]{\includegraphics[width=0.48\textwidth]{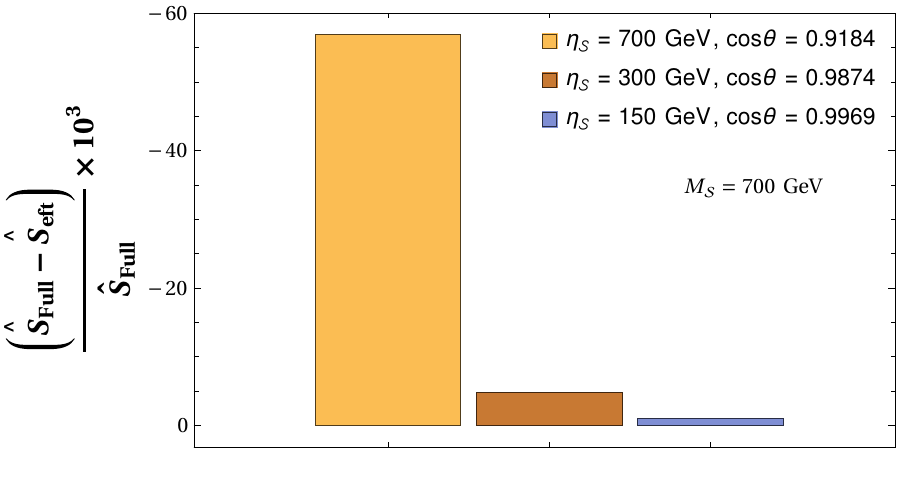}}
     \hfill
     \subfigure[~]{\includegraphics[width=0.48\textwidth]{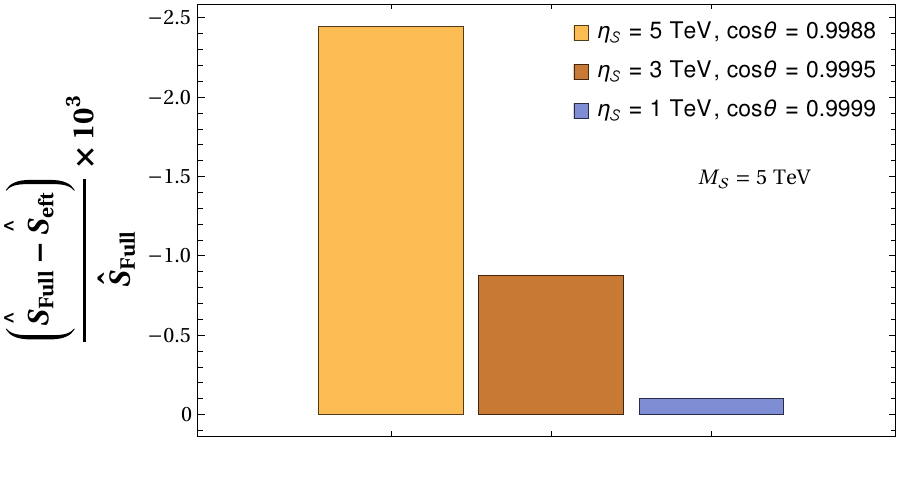}}\hfill
     \parbox{0.48\textwidth}{
     \vspace{-2cm}\caption{Relative difference between full theory and the EFT computation for $\widehat{S}$ parameter at different heavy field mass scales. In this case, we have performed the EFT expansion to accommodate contributions from up to dimension eight effective operators including the running of all contributory operators as well. The mass scales are chosen to be (a) 700 GeV, (b) 1 TeV, and (c) 5 TeV. The values for $\eta_S$ are chosen such that they satisfy the unitarity bounds.}
     \label{fig:deltaS_for_constant_eta}}
\end{figure}
We investigate the departure of the truncated-EFT computation at dimension six from the full theory calculations and the role that the Higgs mixing plays in matching these two. The mixing can be expressed as a function of the trilinear coupling $\eta_S$ and the heavy cut-off scale $M_{\mathcal{S}}$, for allowed $\eta_S$ values, the decoupling can be quantified through the difference of the two theories. In Fig.~\ref{fig:eta_vs_M}, we show the lines for the constant mixing angles that allow a single $\eta_S$ value for each choice of the cut-off. We also impose the constraint from the perturbative unitarity that rules out a specific region in the $\eta_S-M_{\mathcal{S}}$ plane, in turn putting a lower-bound for the mixing for each value of the cut-off $M_{\mathcal{S}}$, that can be seen in Fig.~\ref{fig:costheta_vs_M}.

Intuitively, adding higher and higher order terms in EFT expansion would take the EFT closer to the full theory. This concept is illustrated through the $\widehat{T}$ parameter in Fig.~\ref{fig:comparison_of_different_terms_deltaT}. Here, we consider two different types of contributions. Firstly, the leading order terms in the expansion, i.e., $d_6$ ones. Then, we add $d_8$ contributions including the $(d_6)^2$ ones. In passing, we want to highlight that though the $d_8$ term adds positively to the difference between the full theory and EFT, the further addition of $(d_6)^2$, the equivalent of $d_8$ ones,  allows us to capture the complete contribution at $d_8$. We note that the correction to $\Delta \widehat{T}$ and $\Delta \widehat{S}$ due to the dimension eight inclusion is of the order of the deviations. Thus, dimension eight interactions may be crucial to bringing EFT predictions close to the full theory for given measured constraints. It is important to highlight here that incorporating the $\Delta_\text{NLO}$ contribution leads to a  $7.5\%$ deviation between $m_S$ and $M_\mathcal{S}$ at $700$ GeV. This is the highest deviation achievable when accounting for $\Delta_{\text{NLO}}$ within the parameter range of our choice. At 1 TeV and 5 TeV, we observe deviations of only $3.2\%$ and $0.1\%$, respectively. Ultimately, it reflects that going to higher order in EFT expansion reduces the gap between full theory and EFT, especially for a relatively large mixing. We perform similar analyses for 
$\widehat{S}$ parameter in Fig.~\ref{fig:comparison_of_different_terms_deltaS}. We draw a similar conclusion as the previous one, and that makes our conclusion more generic. 

In Fig.~\ref{fig:deltaT_for_constant_eta}, we have calculated the relative difference between full theory and EFT while calculating the $\widehat{T}$ parameter for three different heavy mass scales. In this case, we have performed the EFT expansion to accommodate contributions up to dimension eight effective operators including the running of all contributory operators as well. In each subfigure, we have shown that if we lower the value of $\eta_S$ for a fixed mass, the value of $\cos \theta$ increases. As the $\cos \theta$ reaches unity, the full theory and EFT are in excellent agreement, which is expected as the new physics contribution vanishes. It is also evident that for a fixed $\eta_S$, once we go for higher masses the difference also decreases. This illustrates the interplay among the coupling $\eta_S$, the heavy mass scale $M_\mathcal{S}$, and mixing parameter $\cos \theta$. One can tune the value of these parameters so that EFT can be a good explanation for the full theory. Doing the same kind of investigation for the $\widehat{S}$ parameter in Fig.~\ref{fig:deltaS_for_constant_eta} further emphasises the idea.

\section{Summary and Conclusions}
\label{sec:conc}

Effective Field Theory is a powerful tool to look for deviations from the SM expectation in a theoretically well-motivated way. In a modern sense, it enables us to extend good quantum field theoretic properties to generic departures from the SM interactions, with potential relevance for UV complete scenarios depending on the accuracy with which constraints can be formulated. Along these lines, a set of particularly well-motivated observables are the oblique corrections as a subset of relevant electroweak corrections. 

In this paper, we have looked into a few queries related to the decoupling limit of a UV theory or in other words the validation of an EFT truncation. We have worked with two oblique parameters $S$ and $T$ as example observables. First, we have estimated these two observables from a full theory perspective and parameterised them in terms of the phyiscal mass of heavy scalar ($M_S$), trilinear coupling ($\eta_S$), and scalar mixing angle ($\cos \theta$). The unitarity limits the choice of trilinear coupling for a given mass of heavy scalar and that in turn helps us to estimate the allowed maximum values of $\cos \theta$. We know that in the alignment limit, i.e. $\eta_S \to\,0$, and decoupling limit, i.e., $M_S \to \infty$  the EFT resembles the full theory, and that is the ideal situation when EFT is valid. We have argued, in this work, that one can approximate a full theory by its effective one even when these parameters are chosen to be far from their ideal values. To establish our claim, we have chosen a few benchmark values (BVs) of these parameters and computed $S$ and $T$ assuming the full theory. Then, we integrated out the heavy scalar and computed the effective operators up to dimension eight that affect these observables. From the EFT perspective, we have further computed $S$ and $T$  and noted the following contributions - from (i)  $d_6$, (ii) $d_6$ as well as $d_8$ operators including the effects from running from contributory effective operators too. We have also incorporated the effects due to the running of $d_6$ operators. We have systematically highlighted, for different BVs,  that while estimating these two observables, the higher dimensional effective operators play a crucial role especially when $M_S$ is not too large. As Higgs boson mixing is a feature in almost all BSM theories with a non-minimal Higgs sector, our analysis shows the necessity to go beyond dimension six interactions when data is very precise or when we want to inform a potential UV scenario accurately.


\section*{Acknowledgements}
 C.E. is supported by the STFC under grant ST/T000945/1, by the Leverhulme Trust under grant RPG-2021-031, and the IPPP Associateship Scheme. M.S. is supported by the STFC under grant ST/P001246/1. W.N. is funded by a University of Glasgow College of Science and Engineering Scholarship.
	
 \appendix
 \section{Gauge boson two-point functions}

\subsection{Modification due to a singlet scalar extension}
\label{app:fullsing}
We note down the modifications to the gauge boson two-point functions due to the presence of a new heavier scalar degree of freedom. Here, only the contributions from the scalar-involved diagrams are presented.
The BSM contribution to the two-point functions (in Feynman gauge) are then \cite{Hahn:2000kx}
\begin{eqnarray}
    \Pi_{ZZ}(p^2) &=&\left[-\frac{M_W^4 B_0\left(p^2,M_h^2,M_Z^2\right)}{4 \pi^2 c_W^4 v^2}+\frac{M_W^2 B_{00}\left(p^2,M_h^2,M_Z^2\right)}{4 \pi ^2 c_W^2
   v^2}-\frac{M_h^2 M_W^2 \left(1-\log
   \left(\frac{M_h^2}{\mu^2}\right)\right)}{16 \pi ^2 c_W^2 v^2}\right]\nonumber\\
   &&\cos^2{\theta}+\left[-\frac{M_W^4 B_0\left(p^2,M_{\cal{S}}^2,M_Z^2\right)}{4 \pi^2 c_W^4 v^2}+\frac{M_W^2 B_{00}\left(p^2,M_{\cal{S}}^2,M_Z^2\right)}{4 \pi ^2 c_W^2
   v^2}\nonumber\right.\\
   &&\left.-\frac{M_{\cal{S}}^2 M_W^2 \left(1-\log
   \left(\frac{M_{\cal{S}}^2}{\mu^2}\right)\right)}{16 \pi ^2 c_W^2 v^2}\right]\sin^2{\theta},\\
   \Pi_{WW}(p^2) &=&\left[-\frac{M_W^4 B_0\left(p^2,M_h^2,M_W^2\right)}{4 \pi ^2
   v^2}+\frac{M_W^2 B_{00}\left(p^2,M_h^2,M_W^2\right)}{4 \pi ^2
   v^2}-\frac{M_h^2 M_W^2 \left(1-\log
   \left(\frac{M_h^2}{\mu^2}\right)\right)}{16 \pi ^2 v^2}\right]\nonumber\\
   &&\cos^2{\theta}+\left[-\frac{M_W^4 B_0\left(p^2,M_{\cal{S}}^2,M_W^2\right)}{4 \pi ^2
   v^2}+\frac{M_W^2 B_{00}\left(p^2,M_{\cal{S}}^2,M_W^2\right)}{4 \pi ^2
   v^2}\right.\nonumber\\
   &&-\left.\frac{M_{\cal{S}}^2 M_W^2 \left(1-\log
   \left(\frac{M_{\cal{S}}^2}{\mu^2}\right)\right)}{16 \pi ^2 v^2}\right]\sin^2{\theta},\\
   \Pi_{\gamma\gamma}(p^2) &=& \Pi_{\gamma Z} (p^2)=0\,,
\end{eqnarray}
where, the Passarino-Veltman functions~\cite{Passarino:1978jh} (see also~\cite{Denner:1991kt,Denner:2019vbn}) $A_0$, $B_0$ and $B_{00}$ capture the scalar one-loop dynamics (the vev is fixed via $v=2M_Ws_W/e$). We have cross checked these results numerically against previous results~\cite{Bowen:2007ia,Englert:2011yb}.

\subsection{Modification due to the corresponding EFT at tree-level}
\label{app:pi-tree-eft}
We note down the tree-level correction to the gauge boson propagators as shown in Fig.~\ref{fig:tree-eft-correction} due to the presence of effective operators.

 \begin{eqnarray}
     \Pi_{WW}^{(\text{EFT})}(p^2) &=& \frac{g_W^2 \,v^6}{16}\mathcal{C}_{H\mathcal{D},1}^{(8)}-\frac{g_W^2 \,v^6}{16}\mathcal{C}_{H\mathcal{D},2}^{(8)}+p^2 v^4 \,\mathcal{C}_{HW}^{(8)}+2 p^2 v^2
   \mathcal{C}_{HW}^{(6)},\\
     \Pi_{ZZ}^{(\text{EFT})}(p^2) &=& \frac{c_W^2 g_W^2 v^6
   }{16}\mathcal{C}_{H\mathcal{D},1}^{(8)}+\frac{c_W^2 g_W^2 v^6
   }{16}\mathcal{C}_{H\mathcal{D},2}^{(8)}+\frac{c_W^2 g_W^2 v^4
   }{8}\mathcal{C}_{H\mathcal{D}}^{(6)}+c_W^2 p^2 v^4\mathcal{C}_{HW}^{(8)}\nonumber\\
   &&+2
   c_W^2 p^2 v^2\mathcal{C}_{HW}^{(6)}+\frac{c_W s_W
   g_W g_Y  v^6
   }{8}\mathcal{C}_{H\mathcal{D},1}^{(8)}+\frac{c_W s_W
   g_W g_Y  v^6
   }{8}\mathcal{C}_{H\mathcal{D},2}^{(8)}\nonumber\\
   &&+\frac{c_W s_W g_W
   g_Y v^4}{4}\mathcal{C}_{H\mathcal{D}}^{(6)}+p^2 c_W
   s_W v^4\, \mathcal{C}_{HWB}^{(8)}+2 p^2 c_W
   s_W v^2 \mathcal{C}_{{HWB}}^{(6)}\nonumber\\
   &&+\frac{
   s_W^2 g_Y^2 v^6 \mathcal{C}_{H\mathcal{D},1}^{(8)}}{16}+\frac{
   s_W^2 g_Y^2 v^6 \mathcal{C}_{H\mathcal{D},2}^{(8)}}{16}+\frac{s_W^2 g_Y^2  v^4
   \mathcal{C}_{H\mathcal{D}}^{(6)}}{8}\nonumber\\
   &&+p^2 s_W^2 v^4
   \mathcal{C}_{HB}^{(8)}+2
   p^2 s_W^2 v^2 \mathcal{C}_{{HB}}^{(6)},\\
     \Pi_{\gamma\gamma}^{(\text{EFT})}(p^2) &=& \frac{c_W^2 g_Y^2 v^6
   }{16}\mathcal{C}_{H\mathcal{D},1}^{(8)}+\frac{c_W^2 g_Y^2 v^6
   }{16}\mathcal{C}_{H\mathcal{D},2}^{(8)}+\frac{c_W^2 g_Y^2 v^4
   }{8}\mathcal{C}_{H\mathcal{D}}^{(6)}+p^2 c_W^2 v^4
   \mathcal{C}_{HB}^{(8)}\nonumber\\
   &&+2 p^2 c_W^2 v^2 \mathcal{C}_{{HB}}^{(6)}-\frac{c_W
   s_W g_W g_Y v^6}{8}\mathcal{C}_{H\mathcal{D},1}^{(8)}-\frac{c_W
   s_W g_W g_Y v^6}{8}\mathcal{C}_{H\mathcal{D},2}^{(8)}\nonumber\\
   &&-\frac{g_W
   g_Y c_W s_W v^4}{4}\mathcal{C}_{H\mathcal{D}}^{(6)}- p^2
   c_W s_W v^4 \mathcal{C}_{HWB}^{(8)}-2 p^2
    c_W s_W v^2 \mathcal{C}_{HWB}^{(6)}\nonumber\\
    &&+\frac{
   s_W^2 g_W^2 v^6 }{16
   }\mathcal{C}_{H\mathcal{D},1}^{(8)}+\frac{s_W^2 g_W^2 
   v^6}{16} \mathcal{C}_{H\mathcal{D},2}^{(8)}+\frac{s_W^2 g_W^2 
   v^6}{8} \mathcal{C}_{H\mathcal{D}}^{(6)}\nonumber\\
   &&+p^2 s_W^2 v^4
   \mathcal{C}_{HW}^{(8)}+2
   p^2 s_W^2 v^2 \mathcal{C}_{HW}^{(6)},\\
    \Pi_{\gamma Z}^{(\text{EFT})}(p^2) &=&-\frac{c_W^2 g_W g_Y v^6}{16}\mathcal{C}_{H\mathcal{D},1}^{(8)}-\frac{c_W^2 g_W g_Y v^6}{16}\mathcal{C}_{H\mathcal{D},2}^{(8)}-\frac{c_W^2 g_W g_Y
   v^4 }{8}\mathcal{C}_{H\mathcal{D}}^{(6)}\nonumber\\
   &&-\frac{ p^2 c_W^2 v^4
   }{2}\mathcal{C}_{HWB}^{(8)}- p^2 c_W^2 v^2
   \mathcal{C}_{{HWB}}^{(6)}+\frac{c_W  s_W g_W^2
   v^6 }{16}\mathcal{C}_{H\mathcal{D},1}^{(8)}+\frac{c_W  s_W g_W^2
   v^6 }{16}\mathcal{C}_{H\mathcal{D},2}^{(8)}\nonumber\\
   &&+\frac{c_W s_W g_W^2
   v^4 }{8}\mathcal{C}_{H\mathcal{D}}^{(6)}-\frac{c_W s_W g_Y^2
   v^6}{16} \mathcal{C}_{H\mathcal{D},1}^{(8)}-\frac{c_W s_W g_Y^2
   v^6}{16} \mathcal{C}_{H\mathcal{D},2}^{(8)}\nonumber\\
   &&-\frac{c_W s_W g_Y^2
   v^4 }{8}\mathcal{C}_{H\mathcal{D}}^{(6)}- p^2 c_W s_W v^4
   \mathcal{C}_{HB}^{(8)}+ p^2 c_W s_W v^4
   \mathcal{C}_{HW}^{(8)}-2
   p^2 c_W s_W v^2 C_{HB}^{(6)}\nonumber\\
   &&+2 p^2 c_W
   s_W v^2 \mathcal{C}_{HW}^{(6)}+\frac{s_W^2 g_W
   g_Y  v^6 }{16}\mathcal{C}_{H\mathcal{D},1}^{(8)}+\frac{s_W^2 g_W
   g_Y  v^6 }{16}\mathcal{C}_{H\mathcal{D},2}^{(8)}+\frac{s_W^2 g_W
   g_Y  v^4 }{8}\mathcal{C}_{H\mathcal{D}}^{(6)}\nonumber\\
   &&+\frac{p^2
   s_W v^4 }{2
   }\mathcal{C}_{HWB}^{(8)}+p^2 s_W^2 v^2\mathcal{C}_{HWB}^{(6)}.
 \end{eqnarray}
The couplings are given by $g_W=e/s_W,g_Y=e/c_W$.

\section{Unitarity Constraints}
\label{sec:unitarity}
Unitarity provides a suitable tool to gauge whether the matching is indeed for perturbative choices of the UV model parameters. Perturbativity, in one way or another, is implicitly assumed in analysing any collider data and this extends to the electroweak precision constraints as well. To this end, we consider the partial wave constraints that can be derived from longitudinal gauge boson scattering to identify the regions of validity this way. 
The zeroth partial wave relevant for this is given for scattering $i_1\, i_2 \to f_1\, f_2$~(see Ref.~\cite{Jacob:1959at})
\begin{equation}
a_{fi}^0 = \frac{\beta^{1/4}(s,m_{i,1}^2,m_{i,1}^2) \,\beta^{1/4}(s,m_{f,1}^2,m_{f,1}^2)}{32\pi s}\int_{-1}^1\text{d}\cos\theta \, \,{\cal{M}}(\sqrt{s},\cos\theta)\,,
\end{equation}
suppressing factors of $1/\sqrt{2}$ for identical particles in the initial $i$ or final state $f$. $\sqrt{s}$ denotes the centre-of-mass energy, and $\theta$ is the scattering angle in this frame for the $2\to 2$ scattering process described by the amplitude $\sim{\cal{M}}$. Furthermore,
\begin{equation}
\beta(x, y, z) = x^2 + y^2 + z^2 - 2xy - 2yz - 2xz\,.
\end{equation}
such that $\lim_{s\to\infty}\beta^2/s = 1$.
Unitarity of the $S$ matrix then translates for $f=i$ to the  conditions
\begin{equation}
\label{eq:pert}
|\text{Re} \, a_{ii}^0 | \leq \frac{1}{2}\quad \hbox{and} \quad | \text{Im} \, a_{ii}^0 |  \leq 1\,.
\end{equation}
of which we use the first one to analyse constraints on the singlet extension parameter space. We only consider the impact of $\eta_S$; perturbative choices of $k_S,\lambda_S$ will not lead to additional unitarity violation.

There are two principle sources of unitarity violation that we consider.
Firstly the presence of a large coupling $|\eta_S|$ can lead to unitarity violation in $HH$ scattering if the size of the coupling is non-perturbative, e.g. comparable to the cut-off scale $M_{\cal{S}}$ from the EFT perspective although the full theory remains convergent (see Eq.~\eqref{eq:expand}).
Secondly large coupling $|\eta_S|$ implies large mixing $\cos^2\theta<1$ and longitundinal gauge boson scattering can set limits as the light Higgs boson only partially restores unitarity there~\cite{Lee:1977eg}. We include $W^+W^-$ scattering as a representative process to obtain unitarity constraints.
Both effects are calculated using FeynArts and FormCalc~\cite{Hahn:1998yk,Hahn:2000jm,Hahn:2000kx} and included to the results of Sec.~\ref{sec:elwprec} for the full theory. We identify the critical value of unitarity violation via
\begin{equation}
|\text{Re} \, a_{ii}^0 (\eta_S,M_{\cal{S}},M_{h},\sqrt{s}=0.9 \,M_{\cal{S}})| = {1\over 2},
\end{equation}
which determines the $\eta_S,M_{\cal{S}}$ contour used throughout this work.

\bibliographystyle{JHEP}
\bibliography{references}
	
\end{document}